# Anomalous and normal dependence of the sound velocity in the liquid Bi-Sb system

Moran Emuna[1], Yaron Greenberg[2], Eyal Yahel[2] and Guy Makov[1]

[1]. Materials Engineering Dept., Ben Gurion University of the Negev, Beer Sheva 84105, Israel.

[2]. Physics Dept., NRCN, P.O. Box 9001, Beer Sheva  84190, Israel.

## Abstract

The sound velocity in selected liquid alloys of the isomorphous Bi-Sb system was measured as a function of temperature to a high accuracy of ±0.2%. The sound velocity temperature coefficient, $d\ln c/dT$, at the liquidus is found to vary non-monotonously as a function of alloy composition, with the transition from normal to anomalous temperature dependence occurring at a composition of approximately $Bi_{35}Sb_{65}$. Beyond this composition up to approximately $Bi_{10}Sb_{90}$ the sound velocity is found to be temperature independent over a wide range. The deviation of the sound velocity from that expected in an ideal solution is found to be dominated by a sub-regular interaction. The present measurements allow the determination of the pressure dependence of the sub-regular solution interaction parameters and are found to be consistent with high pressure studies of the phase diagram in this system. The sound velocity is shown to be an effective measure of the pressure dependence of the alloy interactions.

## Keywords

# 1. Introduction

The physical properties, such as density, heat capacity and sound velocity of some liquid elements have been found to vary in a complex manner across pressure-temperature phase space [1]. Observed behaviors have included non-monotonous temperature and pressure dependences as well as discontinuities which have been taken to indicate the possible existence of liquid-liquid transitions. Experimental determination of these physical properties is challenging due to the high temperatures necessary and the enhanced reactivity of the liquids under these conditions [2].

The sound velocity of liquid metals is an important thermo-physical property, related to the adiabatic compressibility and the dynamic response of liquids, providing a window into the high pressure behavior of the material system. Whereas relatively extensive studies have been performed on many pure liquid elements [3], only a few studies have been made of the physical properties of the multitude of alloy systems [4-9,15,16], despite the relevance of the latter in applications and the possibility of encountering novel phenomenology due to the additional degree of freedom of the composition. Of the binary alloy systems, arguably the simplest is the isomorphous system where the solid phase remains in the same structure at all compositions along the solidus. The thermodynamic theory of such systems at ambient pressure is well known [10], but high pressure modeling of these and other alloy systems is still work in progress [11].

In normal liquid metals, such as Pb, the sound velocity decreases with increasing temperature approximately linearly [12,13]. However, there also exist anomalous liquid metals such as Si, Ge, Sb and Te [3,13] that exhibit non-monotonic temperature dependence of the sound velocity, which first rises to a maximum and then proceeds to decrease. Semi-normal liquids have monotonic, but variable, temperature dependence of the sound velocity, such as Bi [3,12]. Focusing on the pnictides, a progression can be observed down the column from the significant sound velocity anomaly in As where the maximum in the temperature dependence has not yet been determined [14], through anomalous (positive) temperature dependence in Sb with a maximum at approximately $850^0$C [12], to semi-normal (variable) temperature dependence in Bi [12].

Whereas the temperature dependence of the sound velocity in liquid alloys with a eutectic phase diagram has been explored in several studies [4-7,15], the sound velocity in isomorphous alloys has apparently been studied only in the Se-Te system [9,16]. Se-rich liquid alloys exhibit strong semiconducting behavior at low temperatures. With increasing Te concentration or increasing temperature the semiconducting state changes to be metallic [9], i.e., there is a significant change in the electronic nature of the system.

The isomorphous Bi-Sb system exhibits complete solubility of the two components at temperatures below the solidus, as depicted in Fig.1 [17]. At low temperatures, the binary solid solution decomposes into two phases, typical of a repulsive interaction between the two components [18]. Careful modeling of the thermodynamics of this system indicates that the deviation from ideal solution behavior is best described by a sub-regular solution model [18].

Hence, the liquid Bi-Sb system is an attractive model to study the physical nature of isomorphous systems. An early study of the Bi-Sb system by ultrasonic techniques, carried out three decades ago, indicated possible complex dependence of the sound velocity on the system composition [19]. However, the sound velocities were measured only in the vicinity of the liquidus temperature, which varies strongly with composition. Furthermore, the accuracy of the measurements was relatively low, leading to large uncertainties in determining the temperature coefficients.

Only a few studies have been published on other physical properties of the liquid Bi-Sb system. Amongst these are resistivity measurements [20], DSC (Differential Scanning Calorimetry) measurements [21] and viscosity measurements [22] all indicating a possible irreversible liquid-liquid phase transition in the vicinity of $1000^0$C. These indications hint at the complex nature of the Bi-Sb system.

In the present work we report on highly accurate measurements of the sound velocity in the system Bi-Sb as a function of temperature and composition. These measurements are analyzed within a thermodynamic model of the liquid and found to deviate from ideal solution behavior. The deviation from ideality is related to the pressure dependence of the molar volume and is found to correspond to a sub-regular solution model of the interaction between Bi and Sb. The effect on the phase diagram of the Bi-Sb system as a function of pressure is discussed.

## 2. Experimental

### 2.1. Measurement technique

The velocity of sound was measured by the pulse-echo technique, wherein acoustic pulses generated from the transducer, travel through a buffer rod (wave guide) into the liquid and are reflected back to the rod and the transducer. (Fig.2). The velocity of the acoustic waves ($c$) is calculated from the ratio of the distance between the bottom of the rod and the crucible ($l$) to the time interval required for the transmitted wave ($t$) to travel that distance, $c = l/t$. Since the distance between the buffer rod and the crucible is not known accurately enough, we modified the pulse-echo technique [12] and measured the sound velocity by displacing the buffer rod by a known distance, $\Delta X$ (Fig.2). This displacement leads to a change of $2\Delta X$ in the path of the wave and $\Delta t$ in the time interval required for the wave to travel through the liquid, hence, the sound velocity is $c = 2\Delta X / \Delta t$. Hence, reducing the error in determining the distance the acoustic wave travels through the liquid metal.

The experimental setup illustrated in Fig.2, consists of an alumina ($Al_2O_3$) crucible containing the liquid sample, positioned in a furnace. A piezoelectric transducer is located on top of the buffer rod.

## 2.2. Experimental procedure

To obtain good quality acoustic signals, it is essential that the buffer rod is parallel to the bottom of the alumina crucible. This was achieved by optimizing the setup through measuring the signal in ethanol at room temperature. Two thermocouples were positioned as shown in Fig.3. The sample's thermocouple, located in the crucible wall allows measurement of the temperature while avoiding interaction between the thermocouple and the liquid metal. The distance of this thermocouple from the liquid sample was less than 1 mm along the entire height of the crucible. Such a setup reduces the systematic error in determining the sample's temperature. The other thermocouple determines the furnace temperature.

To obtain wetting of the buffer rod by the liquid metal, it was immersed in the melt only at high temperatures (approximately $1000^0C$) obtained after heating at a typical rate of 10°C/min. Measurements were taken upon cooling the sample to the liquidus. The temperature in the sample varied within $\pm 3^0C$ during the measurement period. Over the entire experiment the sample was retained under a protective flow of argon gas to reduce oxidation of the liquid surface.

## 2.3. Error analysis

The errors in determining the sound velocity originate from errors in determining the distance and the time interval required for the sound waves to travel through the liquid. The nominal acoustic path was 1mm for all measurements and the error in the acoustic path $\Delta X$, attributed mainly to the linear motor, is 1 µm. The oscilloscope time base resolution is 2 ns and the typical travel time along the acoustic path is 1 µs. Thus, the total error in determining the sound velocity resulting from the distance and time is estimated to be 0.2%. Errors, resulting from thermal expansions of the structural materials are negligible in the modified pulse-echo method.

## 3. Results

## 3.1. The temperature dependence of sound velocity in liquid Bi-Sb system

The temperature dependence of the ultrasonic sound velocity in liquid Bi and in the liquid alloys $Bi_{70}Sb_{30}$, $Bi_{53}Sb_{47}$, $Bi_{35}Sb_{65}$ and $Bi_{13}Sb_{87}$ was measured and the results are presented in Fig.4. For each sample we obtained several sets of measurements which were found to agree within the experimental error. The arrows on each curve in Fig. 4, indicate the liquidus temperature below which the sample is in the mixed-phase zone (Fig. 1).

The sound velocity in liquid bismuth was measured from $825^0$C down to the liquidus temperature, $271^{o}$C. The sound velocity near the melting point is found to be $1651 \pm 3$ m/s, in excellent agreement with the previously measured value, $1650 \pm 4$ m/s [12]. The sound velocity temperature coefficient, represented by the slope $d\ln c/dT$ at the melting point, is approximately zero, in agreement with Refs.[12] and [23] but not with Ref. [19]. The uncertainty in the temperature coefficient was estimated to be $\pm 1.5*10^{-5}K^{-1}$. The sound velocity decreases monotonously but at an increasing rate with increasing temperature.

The sound velocity in liquid $Bi_{70}Sb_{30}$ was measured from 868°C down to the liquidus temperature (410°C). The sound velocity near the liquidus is found to be 1697 ± 3 m/s. The temperature coefficient at the liquidus temperature is $d\ln c/dT = -5.9*10^{-5}K^{-1}$ with an uncertainty of $\pm 1.8*10^{-5}K^{-1}$.

The sound velocity in liquid $Bi_{53}Sb_{47}$ was measured from 790°C down to the liquidus temperature (475°C). The sound velocity near the liquidus is 1749 ± 3 m/s. The temperature coefficient at the liquidus temperature is $d\ln c/dT = -4.6*10^{-5} \pm 2.3*10^{-5}K^{-1}$.

The sound velocity in liquid $Bi_{35}Sb_{65}$ was measured from 908°C down to the liquidus temperature (537°C). The sound velocity near the liquidus (537°C) is found to be 1784 ± 3 m/s. The temperature coefficient, $d\ln c/dT$ is approximately zero over a temperature range of 124°C, extending from 537°C to 661°C, with an uncertainty of $\pm 1.35*10^{-5}K^{-1}$.

Finally, the sound velocity in liquid $Bi_{13}Sb_{87}$ was measured from 830°C down to the liquidus temperature (607°C). At this composition we see that the sound velocity increases at the liquidus with a temperature coefficient of $d\ln c/dT = 5.9*10^{-5}K^{-1} \pm 2.8*10^{-5}K^{-1}$. The sound velocity forms a broad maximum with a value of ca. 1870 m/s at approximately 690°C and proceed to decline slowly at higher temperatures.

### 3.2. The temperature dependence of the sound velocity

The derivative of the sound velocity with respect to temperature ($d\ln c/dT$) at the liquidus of the Bi-Sb system, as a function of composition is shown in Fig.5 based on the results of the present measurements and previous studies [12,19]. It is found to vary non-monotonously as a function of composition. The anomalous increase of sound velocity with temperature, found in pure Sb, weakens with the addition of Bi and the temperature coefficient becomes zero at approximately 65at% of Sb which signals a transition from anomalous (positive) to normal (negative) behavior. As the fraction of Bi is further increased the normal decrease of sound velocity with temperature strengthens until maximum (negative) temperature dependence is obtained. Further increase of the fraction of Bi causes the temperature coefficient to become smaller again and it is approximately zero for pure Bi.

Fig.6 shows the isothermal composition dependence of the sound velocity at selected temperatures based on the present work and the results of Ref.[12]. At all temperatures, the sound velocity increases monotonously with composition as the fraction of antimony is increased. However, at high temperatures the rate of change increases as the temperature is increase a phenomenon which also occurs in Bi rich compositions at low temperatures. This leads to a change in the temperature dependence as a function of composition. We note that the dependence of the sound velocity on temperature is very weak over a range of ca. 350 degrees for compositions in the range of 65at% to 90at%Sb.

## 4. Discussion

### 4.1. Temperature dependence on the velocity of sound in the Bi-Sb system

In liquid Bi, the sound velocity decreases with temperature as observed in most liquid metals [12]. In contrast, the sound velocity in liquid Sb increases with temperature up to a maximum at approximately $850^oC$, following which the temperature dependence returns to normal and the sound velocity decreases with temperature [3]. In the present study we have explored the effect of alloy composition on the sound velocity in the binary system Bi-Sb at temperatures below $1000^0C$, where a possible change in the liquid structure has been indicated [20-22]. The sound velocity as a function of temperature in liquid Bi, $Bi_{70}Sb_{30}$, $Bi_{53}Sb_{47}$, $Bi_{35}Sb_{65}$ and $Bi_{13}Sb_{87}$ was measured by a modified pulse-echo technique to high accuracy, resulting in an experimental error of ca. 0.2%. The temperature coefficients were obtained with a typical uncertainty of ca.$\pm 1.3$-$2.8*10^{-5}K^{-1}$ as compared to $\pm 4$-$8.9*10^{-5}K^{-1}$ in earlier measurements. From these results, we find that the temperature dependence of the sound velocity changes from anomalous (positive) behavior in the antimony-rich compositions to normal (negative) behavior in bismuth-rich compositions. The crossover from anomalous to normal behavior occurs at ca. 65%Sb. In the present section we analyze these results within a thermodynamic model of the solution.

**4.2. Empirical deviation from ideal solution**

The velocity of sound in liquids is related to the adiabatic compressibility ($K_s$) by:

$$4.1) \quad \frac{1}{c_s^2} = \rho K_s = -\frac{M}{V^2}\left(\frac{\partial V}{\partial P}\right)_s.$$

where $\rho$ is the mass density, $M$ is the molar mass and $V$ the molar volume. We define the volume of a binary system $AB$ as a function of composition by:

$$4.2) \quad V = X_A V_A + X_B V_B + \delta V \equiv V_{id} + \delta V,$$

where $X_A$ and $X_B$ are the mole fractions of each component and $V_A$ and $V_B$ are the molar volume of the two components (e.g., Bi, Sb). $\delta V$, represents the deviation of the molar volume from that of an ideal solution. $M = X_A M_A + X_B M_B$, is the molar mass for the solution.

The sound velocity expected in an ideal solution may be defined as:

4.3) $$\frac{1}{c_{s,id}^2} = -\frac{M}{V_{id}^2}\left(\frac{\partial V_{id}}{\partial P}\right)_s = -\frac{M}{V_{id}^2}\left(X_A\left(\frac{\partial V_A}{\partial P}\right)_s + X_B\left(\frac{\partial V_B}{\partial P}\right)_s\right)$$

Fig.7 shows the deviation of the measured sound velocity, from that expected in an ideal solution, $(c_{id}/c_s)^2 - 1$ for Bi, Bi$_{70}$Sb$_{30}$, Bi$_{53}$Sb$_{47}$, Bi$_{35}$S$_{b65}$, Bi$_{13}$Sb$_{87}$ and Sb at several temperatures. It is seen, that the deviation of the experimental sound velocity from that of an ideal solution is asymmetric in composition and decreases to a minimum at Sb-rich compositions of ca.70%Sb, consistent with a sub-regular interaction between the alloy components [18].

### 4.3. Deviation from ideality

To explore the origin of the deviation of the sound velocity from that of an ideal solution, we express $c_s$ as a function of the volume, $V = V_{id} + \delta V$. Employing Eq.4.3 we obtain:

4.4) $$\frac{1}{c_s^2} = -\frac{M\left(X_A\left(\frac{\partial V_A}{\partial P}\right)_s + X_B\left(\frac{\partial V_B}{\partial P}\right)_s\right) + \frac{\partial \delta V}{\partial P}}{(X_A V_A + X_B V_B + \delta V)^2}$$

Expanding to first order in $\delta V$, we obtain:

4.5) $$\left(\frac{c_{id}}{c_s}\right)^2 - 1 = -\frac{2\delta V}{V_{id}} + \frac{\frac{\partial \delta V}{\partial P}}{X_A \frac{1}{c_{s,A}^2}\left(-\frac{V_A^2}{M_A}\right) + X_B \frac{1}{c_{s,B}^2}\left(-\frac{V_B^2}{M_B}\right)} + O(\delta V)^2$$

We find that the deviation of the sound velocity from an ideal behavior is composed of two terms. The first term on the r.h.s in Eq. (4.5) depends on $\delta V$ and reflects the contribution due to the deviation of the molar volume from ideality and contributes negatively to the deviation of the sound velocity from an ideal solution. The second term, reflects the contribution of the pressure dependence of the deviation from the ideal molar volume, ($\partial\delta V/\partial P$) and contributes positively to the above mentioned deviation.

In order to ascertain which of these terms is dominant, we can estimate the contribution of the first term from measurements of the molar volume as a function of concentration. To the best of our knowledge, there are no data for the molar volumes of liquid Bi-Sb alloys. Instead, $\delta V/V_{id}$ was estimated from measurements of the lattice parameter obtained for Bi-Sb solid system [24,25], under the assumption that the repulsive interactions in the alloy are essentially state independent. The values obtained for $\delta V/V_{id}$ which is the first term on the r.h.s of Eq. 4.5, are typically less than 0.25%, an order of magnitude less than the values obtained empirically for the deviation of the sound velocity from ideality (l.h.s of Eq. 4.5) and shown in Fig.7. Therefore, it is seen that the deviation in volume contributes negligibly to the relative deviation of the sound velocity which is of the order of several percent. Hence, the pressure dependence of the deviation of the molar volume, $\partial\delta V/\partial P$, cannot be neglected. This deviation is asymmetric with respect to composition and is represented by an asymmetric fit to the data, indicating the importance of the sub-regular contributions to the free energy and its derivative at all temperatures.

### 4.4. The pressure effects

The asymmetry of the composition dependence of the deviation from ideality indicates that the system may be considered as a sub-regular solution in agreement with phase diagram modeling [18]. The Gibbs free energy of a non-ideal sub-regular solution is [26]:

4.6)    $G = X_A G_A + X_B G_B + RT(X_A \ln X_A + X_B \ln X_B) + J_0 X_A X_B + J_1 X_A X_B (X_A - X_B)$

where $J_0$ and $J_1$ are the regular and sub-regular solution interaction coefficients. Within this model, the molar volume $V$ is given by,

$$4.7) \quad V = X_A V_A + X_B V_B + \frac{\partial J_0}{\partial P} X_A X_B + \frac{\partial J_1}{\partial P} X_A X_B (X_A - X_B)$$

Comparing to Eq. 4.2, the deviation of the molar volume is given by:

$$4.8) \quad \frac{\partial J_0}{\partial P} X_A X_B + \frac{\partial J_1}{\partial P} X_A X_B (X_A - X_B) = \delta V$$

The pressure dependence of the interaction coefficients $J_0$, $J_1$ can be expanded to 2nd second order, where the second derivative of the interaction parameters with respect to pressure related to the deviation in the volume, $\delta V$ by 4.8 and by:

$$4.9) \quad \frac{\partial^2 J_0}{\partial P^2} X_A X_B + \frac{\partial^2 J_1}{\partial P^2} X_A X_B (X_A - X_B) = \frac{\partial \delta V}{\partial P}$$

Hence, we can relate the pressure dependence of the interaction parameters $J(P)$ to the measured quantities, $\delta V$ and $\partial \delta V / \partial P$.

As an example, we have calculated the following parameters using Eq.4.8 and Eq.4.9 for two compositions of $Bi_{70}Sb_{30}$ and $Bi_{35}Sb_{65}$ to find:

$$\frac{\partial J_0}{\partial P} = 2*10^{-7} \, J/molPa, \quad \frac{\partial J_1}{\partial P} = 3*10^{-7} \, J/molPa,$$

$$\frac{\partial^2 J_0}{\partial P^2} = -2*10^{-17} \, J/molPa^2, \quad \frac{\partial^2 J_1}{\partial P^2} = 1*10^{-16} \, J/molPa^2.$$

The order of magnitude of these values remains unchanged for different choices of composition. Thus, we see that measurements of the sound velocity in liquid alloys allow the determination of the pressure dependence of the interaction parameters in the alloy.

Based on these results we can estimate the pressure dependence of the critical temperature for phase separation in Bi-Sb solution given by $T_c = J_0/2R$ in the regular solution model. $J_0$, at ambient pressure, is determined from the $T_c$ estimated from the phase diagram [Fig.1] to be 445K. The pressure dependence of the interaction coefficients $J_0$ is given by the relation:

4.10) $\quad J_0(P) = 7400 + 2*10^{-7}P - 10^{-17}P^2$,

where $P$ is in Pa and $J_0$ in J/mole. Hence, the pressure dependence of the critical temperature in K is:

4.11) $\quad T_c(P) = 445 + 10^{-8}P - 6*10^{-19}P^2$

The effect of the pressure up to several GPa on the phase diagram of Bi-Sb can now be examined. The fusion curve of antimony exhibits practically no change up to these pressures but in pure bismuth the melting temperature decreases approximately to 460K at 2GPa [27]. The critical temperature, below which a miscibility gap appears, increases with pressure from 445K at ambient to approximately 464K at 2GPa. As a result, in the composition range of ca.20%-50% Sb, upon heating, the solidus will intersect the miscibility line before complete miscibility is achieved. Hence, upon heating, a "pure" Bi melt will appear before melting of the alloy. This prediction is consistent with high pressure thermal measurements of Bi-Sb system which have been interpreted to indicate melting of pure Bi prior to that of the alloy [28].

## 5. Conclusions

We have studied the temperature dependence of the sound velocity as a function of composition in the isomorphous Bi-Sb system. We have found that the transition from anomalous temperature dependence in Sb to normal temperature dependence in Bi as a function of alloy composition is not monotonous. The sound velocity reflects the underlying, sub-regular, interaction between the alloy components which has been previously identified by thermal measurements. We have found that the pressure dependence of the interaction parameters can be deduced from the measured sound velocities. Furthermore, these thermodynamic parameters derived from the sound velocity measurements at ambient pressure are consistent with high pressure studies of the melting curve and the phase diagram in general. This result indicates that sound velocity studies in liquid alloys can provide important pressure dependent information concerning the phase diagram of these systems in the liquid, and most probably in the solid phases as well.

## Acknowledgements

The authors acknowledge support of the joint Israel CHE-IAEC research fund.

**Figures:**

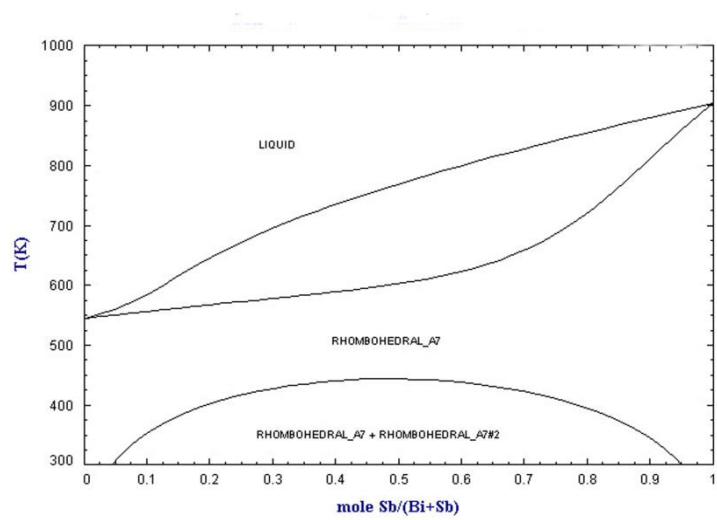

Fig.1. Phase diagram of bismuth-antimony from the SGTE database generated with FactSage [17].

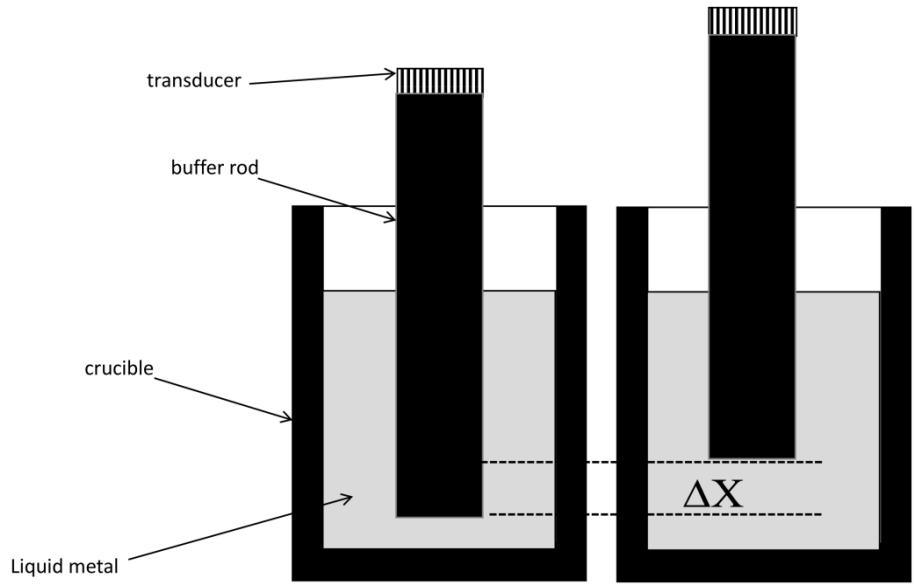

Fig.2. Schematic view of the experimental configuration illustrating the modified pulse-echo technique.

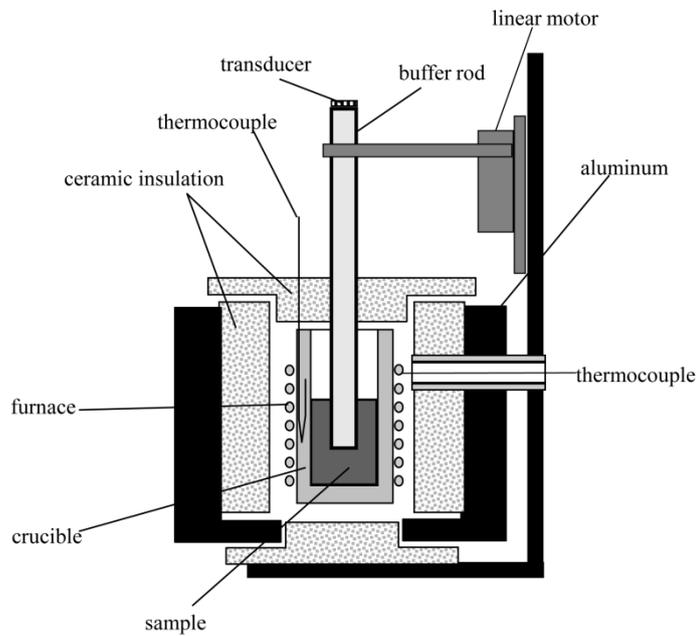

Fig.3. Schematic view of the experimental measurement apparatus.

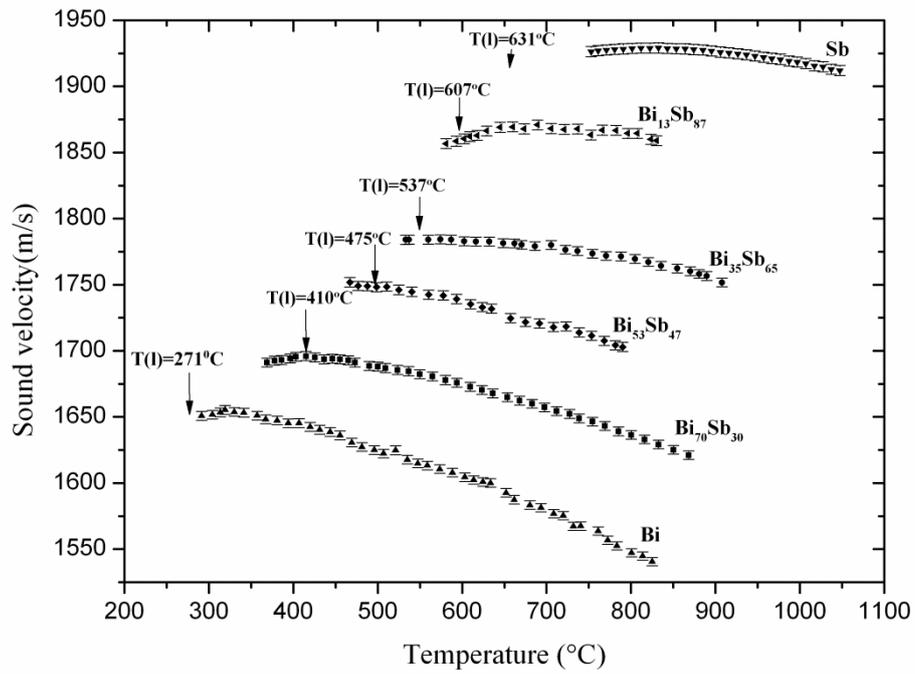

Fig.4. Sound velocity in the liquid Bi-Sb system as a function of temperature at selected alloy compositions.

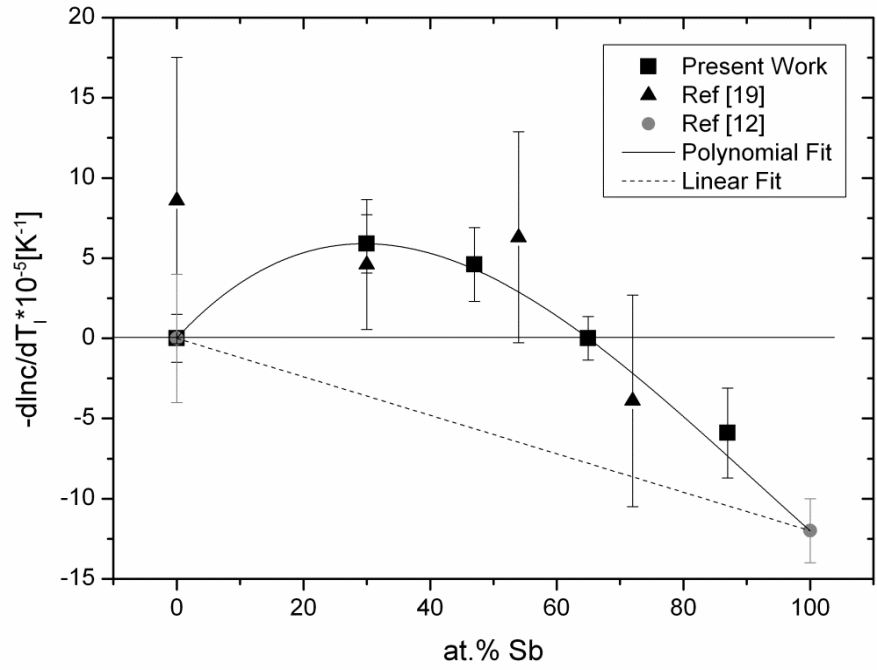

Fig.5. The temperature dependence of the sound velocity at the liquidus temperatures as a function of composition in bismuth-antimony alloys, including results of present work and measurements from previous studies [12,19].

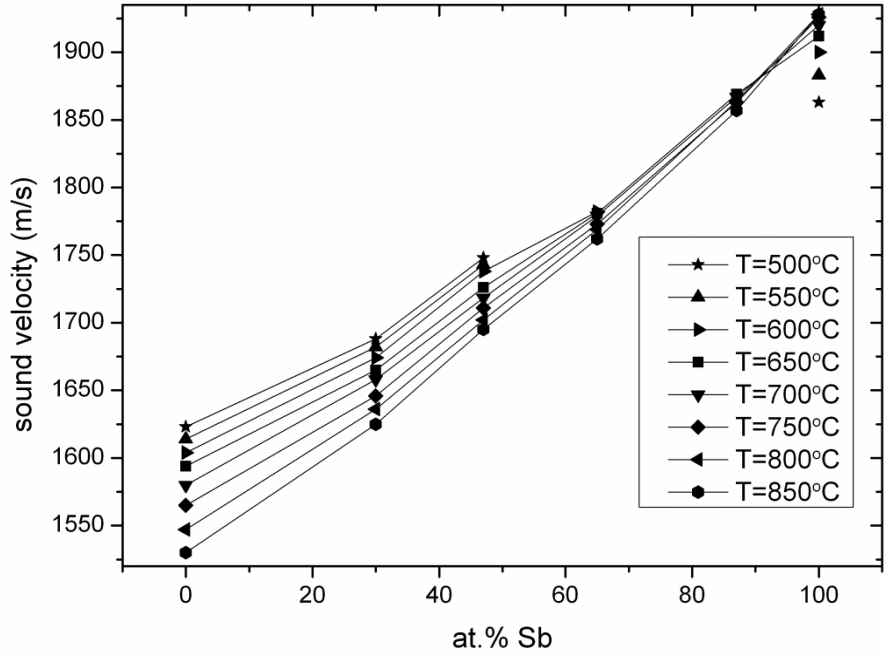

Fig.6. The sound velocity dependence on composition in liquid bismuth-antimony alloys, at selected constant temperatures [12]. Note that the symbols are larger than the error bar

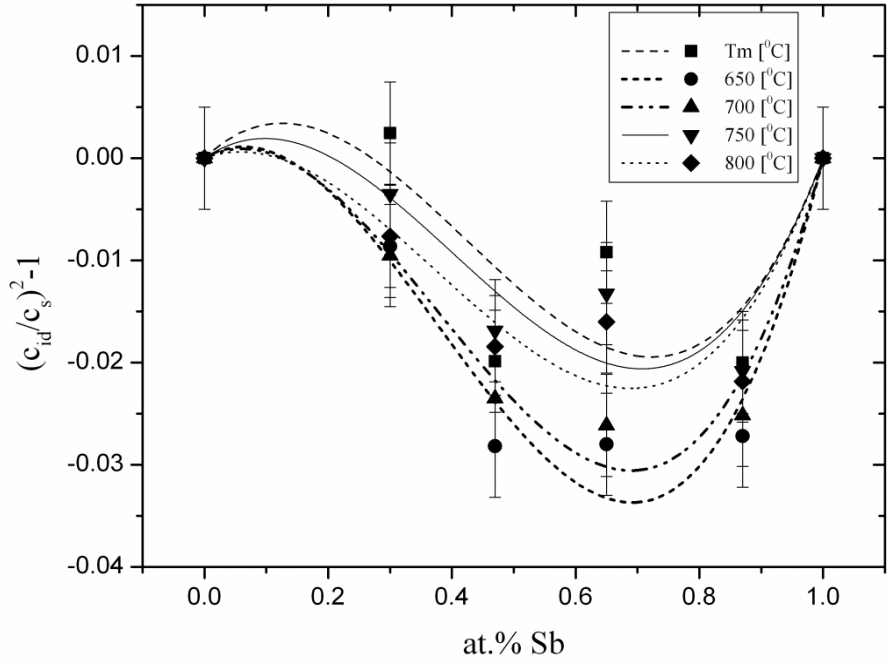

Fig.7. The empirical deviation of the sound velocity from an ideal solution measured as function of antimony compositions at several temperatures. Lines are 3rd order polynomial fits to the data, representing the asymmetric dependence on composition, as reflected in the sub-regular solution model.